\documentclass[twocolumn]{aastex63}

\usepackage{graphicx}	
\usepackage{amsmath}	
\usepackage{amssymb}	
\usepackage{hyperref}
\usepackage{multirow}

\newcommand{\ph}[1]{\phantom{#1}}

\newcommand{\DP}{DIPol-2}
\newcommand{\DUF}{DIPol-UF}
\newcommand{\code}[1]{\texttt{#1}}
\newcommand{\CS}{C\#}

\shorttitle{DIPol-UF: simultaneous three-color ($BVR$) polarimeter with EM CCDs}
\shortauthors{Piirola et al.}

\begin{document}

\title{DIPol-UF: Simultaneous Three-Color ($BVR$) Polarimeter with EM CCDs}

\author[0000-0003-0186-206X]{Vilppu Piirola}
\affiliation{Department of Physics and Astronomy, FI-20014 University of Turku, Finland}

\correspondingauthor{Vilppu Piirola}
\email{piirola@utu.fi}

\author[0000-0001-5563-7840]{Ilia A. Kosenkov}
\affiliation{Department of Physics and Astronomy, FI-20014 University of Turku, Finland}

\author[0000-0002-9353-5164]{Andrei V. Berdyugin}
\affiliation{Department of Physics and Astronomy, FI-20014 University of Turku, Finland}

\author[0000-0002-2238-7416]{Svetlana V. Berdyugina}
\affiliation{Leibniz-Institut f\"{u}r Sonnenphysik, Sch\"{o}neckstr. 6, 79104 Freiburg, Germany}

\author[0000-0002-0983-0049]{Juri Poutanen}
\affiliation{Department of Physics and Astronomy, FI-20014 University of Turku, Finland}

\begin{abstract}
{We describe a new instrument capable of high precision ($10^{-5}$) polarimetric observations simultaneously in three passbands ($BVR$). The instrument utilizes electron-multiplied EM CCD cameras for high efficiency and fast image readout. The key features of \DUF\ are: (i) optical design with high throughput and inherent stability; (ii) great versatility which makes the instrument optimally suitable for observations of bright and faint targets; (iii) control system which allows using the polarimeter remotely. Examples are given of the first results obtained from high signal-to-noise observations of bright nearby stars and of fainter sources such as X-ray binaries in their quiescent states.}
\end{abstract}


\section{Introduction}
\label{sec:introduction}

The development of fast-readout imaging devices like electron-multiplied CCDs (EM CCDs) has opened new possibilities to enhance the efficiency of fast multiple exposure instruments, such as photopolarimeters. Readout noise is suppressed to negligible levels for faint targets. Combined with the idea of grossly defocused images for bright sources, allowing to collect up to $10^8$ photons in one exposure without saturating pixels, which provides the high signal-to-noise ratio (S/N) required for precision polarimetry, has made it possible to construct versatile new instruments for photopolarimetric applications.

With conventional CCDs the previous version of our instrument, the \DP\ polarimeter \citep{Piirola2014},  reaches the precision $10^{-5}$ in about one hour of total telescope time for sufficiently bright stars (photon noise limited). These cameras have image readout and download times 0.7--1.0 s, which multiplied by the number of exposures, typically in the range 512--1024 for one high S/N observation, builds up a significant amount of time lost between the exposures. The EM CCDs have readout times $< 0.02$ s and also the option to co-add frames in the camera memory before download, which further improves the efficiency.

In the present paper we describe a new polarimeter, \DUF, employing three Andor iXon Ultra 897 EM CCDs for simultaneous polarimetry in three passbands ($BVR$) separated by color-selective beam splitters (dichroic mirrors). We also show some first results obtained from high S/N observations of nearby stars, of fainter X-ray binaries containing accreting black holes, and a white dwarf with a debris disk.

\section{Design of the polarimeter}
\label{sec:design}

The design of \DUF\ (Double Image Polarimeter -- Ultra Fast) is based on that of its successful predecessor, \DP\ \citep{Piirola2014, Piirola2020}, which uses a superachromatic half-wave plate (HWP) as the modulator and a plane-parallel calcite plate as the polarizing beam splitter. 
The polarimeter simultaneously records two orthogonally polarized beams from each source in the field of view. 
For the sky background, the polarized components overlap in the focal plane and the sky is not split into polarized components. Hence, sky polarization is effectively optically eliminated at the instrument \citep{Piirola1973}.
Double-image polarimeters suffer no systematic errors that arise in the case of simple, one-image design \citep[see, e.g.,][]{Clarke1965}.
Dichroic beam-splitters enable simultaneous measurements in multiple optical passbands (usually, $B$, $V$, and $R$), which greatly improves efficiency. There is very little internal absorption in the color selective beam splitters.

\subsection{Optical system}
\begin{figure}
    \includegraphics[keepaspectratio, width = 0.95\linewidth]{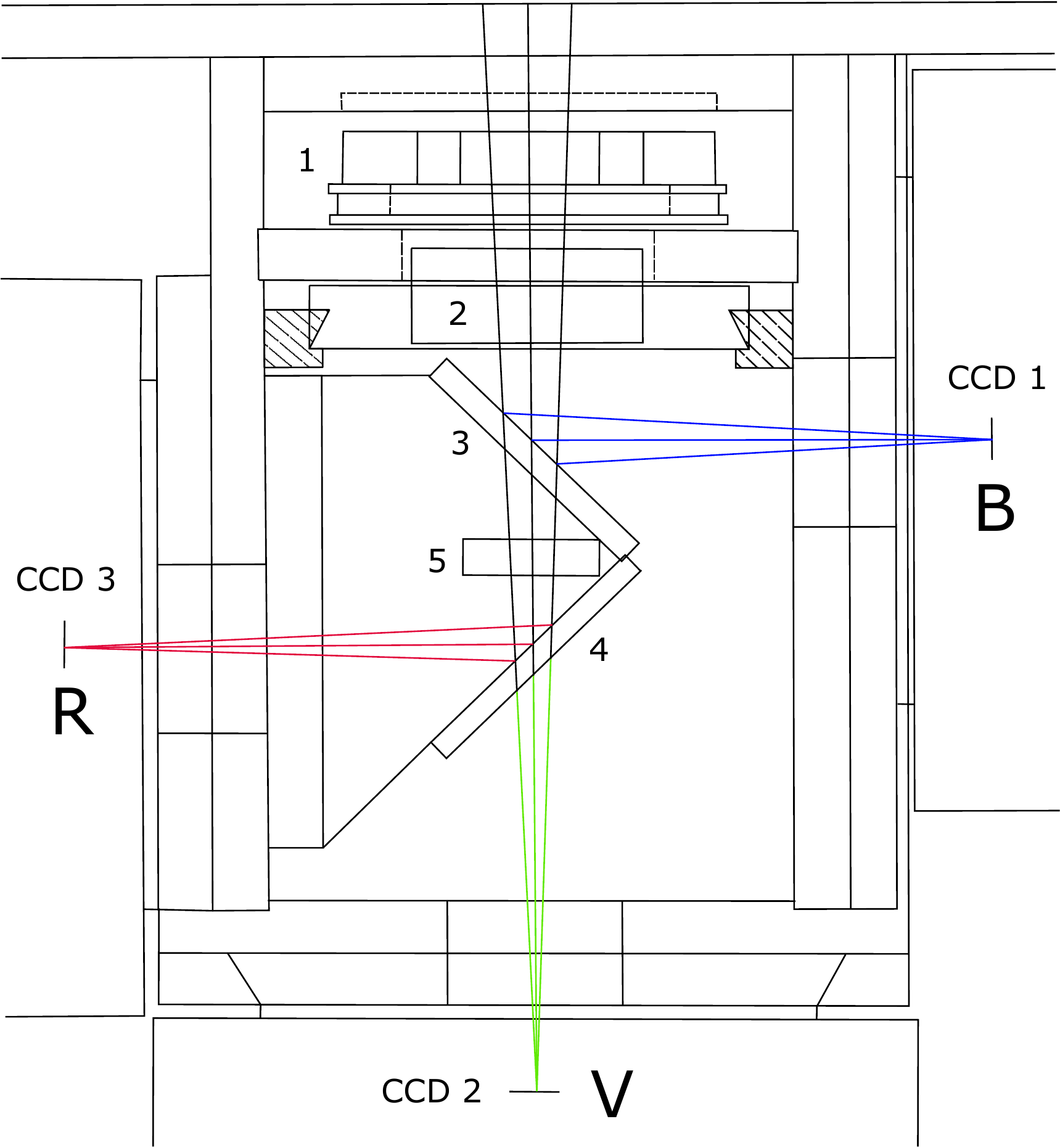}
    \caption{
        Schematic representation of the side view of \DUF. 
        1. Modulator -- a retarder plate (either $\lambda/2$ or $\lambda/4$), rotated by a stepper motor.
        2. Polarization analyzer - a calcite unit, retractable by another stepper motor.
        3. First dichroic beam-splitter (blue reflector).
        4. Second dichroic beam-splitter (red reflector).
        5. Focal extender lens.}
    \label{fig:optics}
\end{figure}

A schematic representation of \DUF\ is shown in Figure~\ref{fig:optics}.
The optical system consists of the following components: a modulator (B. Halle superachromatic HWP), a retractable calcite unit, two dichroic beam-splitters (blue reflector and red reflector), a focal extender lens and three  detectors -- electron-multiplying charge-coupled devices (EM CCD). 
The light enters \DUF\ through the modulator -- a $\lambda/2$ retarder plate for linear and $\lambda/4$ plate (QWP) for circular polarization measurements.
The plate is rotated at discrete positions, 22\fdg5 intervals for linear and 90\degr\ for circular polarimetry, by a stepper motor and remains at fixed angle while an exposure is in progress.

The plane-parallel calcite plate acts as a polarization analyzer and splits the incident radiation into two orthogonally polarized rays (ordinary and extraordinary). The separation of the o- and e-images is proportional to the thickness, $d$, of the plate. With $d$ = 18.0 mm the beam separation is 2 mm, which is convenient for all but the longest focal length telescopes. For example, with the image scale of the 2.5 m Nordic Optical Telescope (NOT) the o- and e-image separation is about 14\farcs6, which is sufficient even in the worst seeing conditions. 

\begin{figure}
    \includegraphics[keepaspectratio, width = 0.95\linewidth]{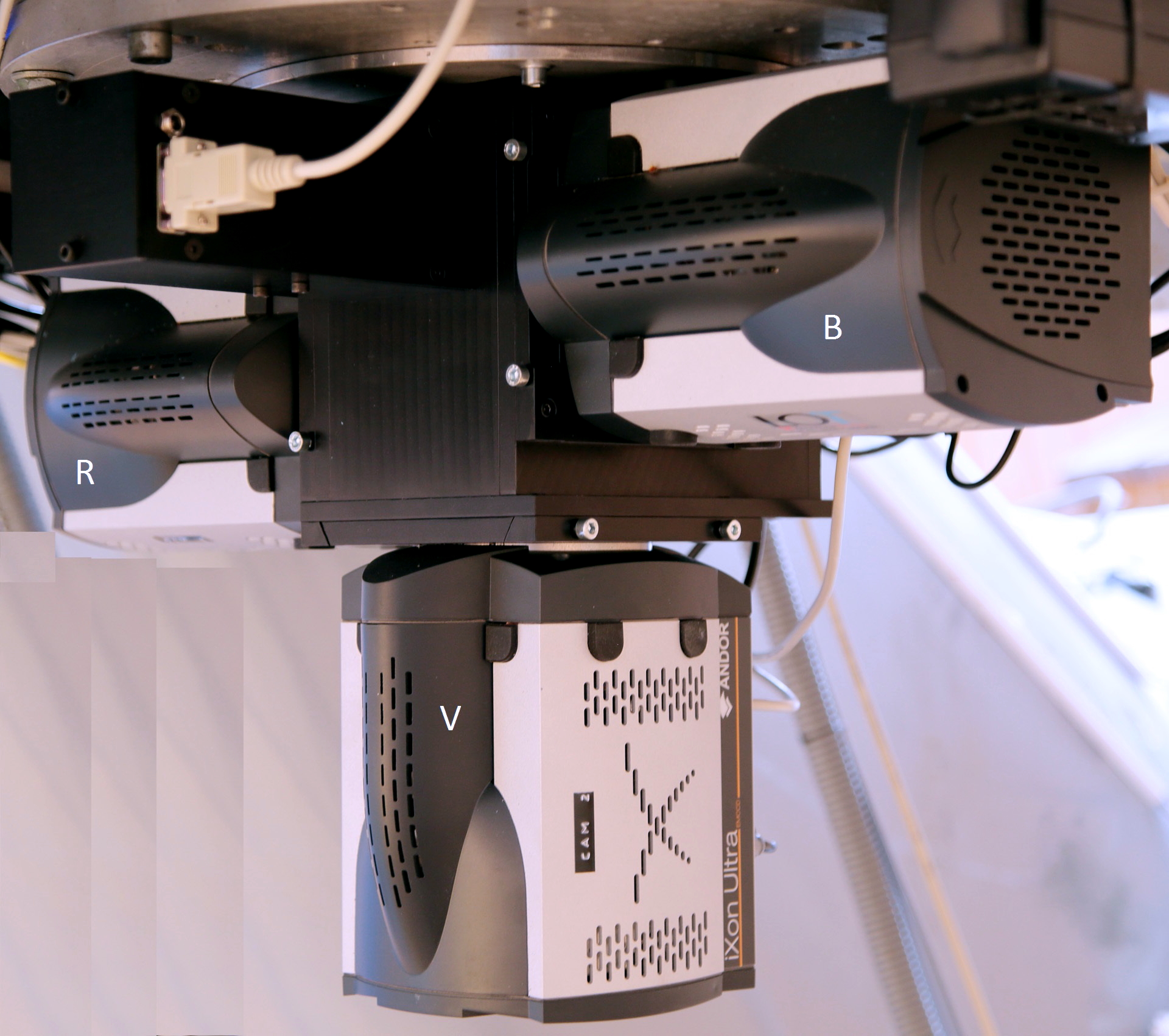}
    \caption{
    \DUF\ polarimeter head shown with the three Andor Ixon Ultra EM CCD cameras for the $B$, $V$, and $R$ passbands.
        }
    \label{fig:duf_photo}
\end{figure}

Two dichroic beam-splitters split the light into the three ($BVR$) spectral passbands, with equivalent wavelengths 450, 545, and 650 nm, and full widths at half maximum (FWHM) 110, 90, and 110 nm, respectively. The images are recorded by three Andor iXon EM CCD cameras (Figure~\ref{fig:duf_photo}). The dichroics are on a thin, $d$ = 1 mm, substrate to minimize image aberrations in the convergent transmitted beam. The calcite plate is oriented so that the o- and e-beam polarization planes are at 45\degr\ from the normal of the dichroic mirrors projected to the focal plane, to avoid large differences in the reflection and transmission coefficients for the two polarized beams.

Due to the physical size of the Andor iXon cameras, and the telescope back focal distance requirements, a weak negative achromatic lens is installed after the first dichroic beam splitter to bring all the three cameras simultaneously into focus. The focal extender lens magnifies the image scale at the $R$ and $V$ cameras by about 1.6. This is useful in the sense that photons are spread onto a larger number of pixels, which reduces the tendency of pixel saturation in these passbands where the sensitivity of the CCDs is the highest.  

The calcite unit can be fully retracted from the optical path with a second stepper motor.
This effectively turns the polarimeter into simultaneous three-color ($BVR$) imaging fast photometer.
 
\subsection{ANDOR EM CCDs}

\DUF\ uses three ultrafast Andor iXon EM CCD cameras.\footnote{\url{https://andor.oxinst.com/products/ixon-emccd-camera-series/ixon-ultra-897}} 
The $V$ and $R$ cameras are identical (iXon Ultra 897-EX model), while for $B$-passband iXon Ultra 897-BV model with the sensitivity optimized for the blue region is chosen.

Unlike most conventional CCD cameras, Andor iXon Ultra EM CCD series offers a \emph{fast frame transfer}. The CCD chip has two areas, one is for the image exposure and another for the temporal image storage. As soon as the exposure is taken, the image is swiftly shifted vertically from the exposure area into the masked storage area from where it is transferred to the camera readout register. This allows a series of short exposures to be taken with permanently open shutter, with no visible image smear.

Each camera has an active area of $512 \times 512$ pixels ($16 \times 16~\mu$m each), and supports readout speed up to 17~MHz.
When used in the EM mode, and with the vertical shift interval of 0.3~$\mu$s, iXon Ultra 897 can achieve readout rate up to 56 full frames per second with no binning.
Fast image readout allows using iXon Ultra camera in `video mode', which produces a continues series of images with minimum delays -- a useful tool for precise pointing and focusing. Camera is equipped with thermoelectric cooling system which provides cooling down to $-80\degr$~C assuming air cooling with ambient temperature of $20\degr$.

High-speed Andor EM CCD cameras have been used in a number of instruments, e.g. RINGO3 \citep{Arnold2012}, GASP \citep{Collin2013}, and GPP \citep{Gisler2016}.
They are well suited for applications, where fast modulation and image readout are required.

iXon Ultra 897 is a highly configurable camera. It features conventional and an electron-multiplying (EM) amplifiers. 
The choice of the amplifier, vertical shift clock speed, pixel readout rate, pre-amplifier gain and EM amplification gain can be made to suit optimally observer needs and the nature of observed target. 
Changing camera parameters can be done on-the-fly via control software. 
Conventional amplifier provides the widest dynamical range and is best fit for polarimetry of bright sources. 
With the optimal vertical shift time of $3.33~\mu$s and 3 MHz horizontal readout rate, the image retrieval time reduces to $\sim0.1$~s.
Faint targets are better recorded in the EM amplifier mode with the EM gain value in the range of 5--20. 
In EM mode, camera can achieve single-photon sensitivity owing to the elimination of the readout noise, which significantly increases the signal to noise ratio (S/N).

The main advantage of \DUF\ in comparison with many other CCD polarimeters follows from its design, which makes this instrument equally efficient for observations of bright as well as faint targets. For the polarimetry of a bright star, the image is defocused and recorded with the conventional amplifier. The degree of defocusing is chosen depending on the stellar brightness and no neutral density filters are needed. For the faintest targets, the image is focused precisely and recorded with the EM amplifier.

\subsection{Polarimetric mode}

\begin{figure*}
\centering 
    \includegraphics[keepaspectratio, width = 0.4\linewidth]{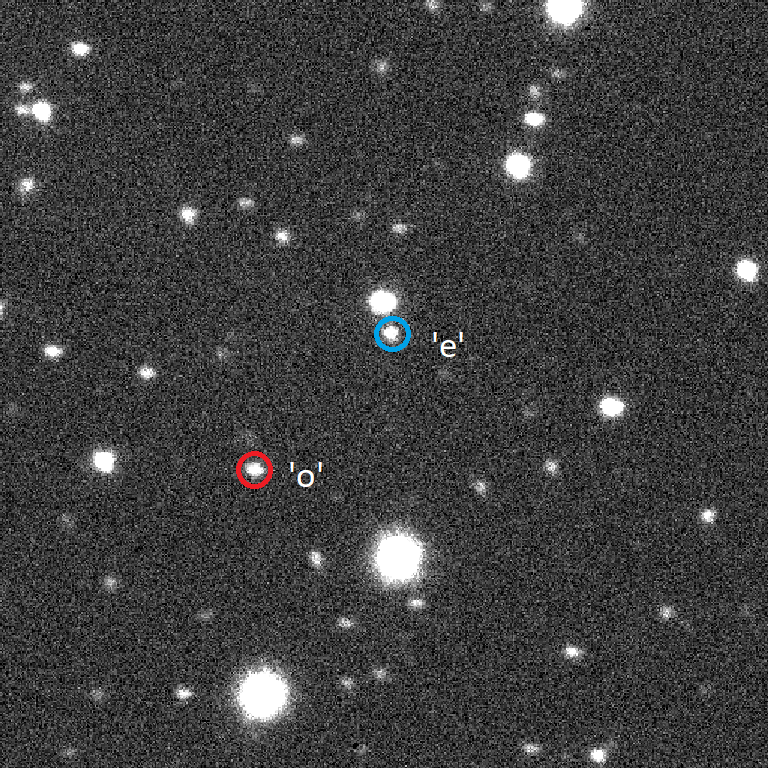}
    \includegraphics[keepaspectratio, width = 0.4\linewidth]{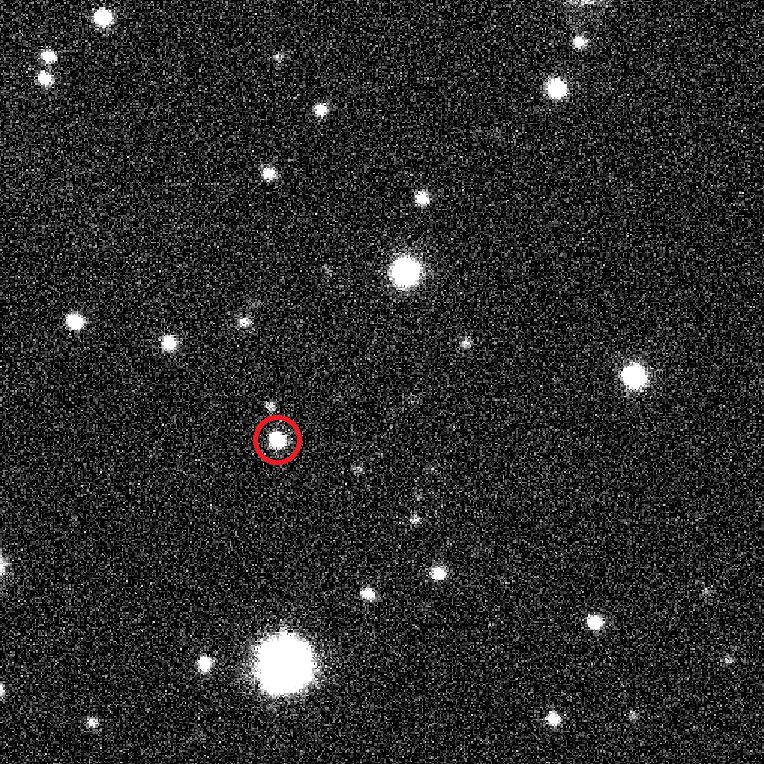}
    \caption{
    Field around black hole X-ray binary MAXI~J1820$+$070 \citep{Denisenko2018,Kawamuro2018} as seen by \DUF\ during its first commissioning run at the NOT in July 2019.
    Left panel: Polarimetric mode. The analyzer produces two images of each star in the field.
    Ordinary and extraordinary rays of MAXI~J1820$+$070 are marked with the red and blue circles with 'o' and 'e' labels, respectively.
    Right panel: Photometric mode. Same region of the sky, but with the analyzer retracted. MAXI~J1820$+$070 is highlighted with the red circle. The field of view at the NOT is $1\farcm0 \times 1\farcm0$ in the $B$-band (0\farcs117 per pixel), and $37\farcs1 \times 37\farcs1$ in the $V$- and $R$-bands (0\farcs0725 per pixel).
}
\label{fig:dipol_maxi_field}
\end{figure*}

Double-image polarimeters require a series of exposures taken at a different modulator position.
One independent measurement of linear polarization is computed from 4 sequential images (e.g., at HWP angles 0\degr, 22\fdg5, 45\degr, and 67\fdg5). One independent measurement of circular polarization requires two images with the QWP angle offset by 90\degr, i.e. at 45\degr\ and 135\degr).
One linear polarization measurement cycle consists of 16 images recorded simultaneously by three CCDs for one full rotation of the HWP modulator ($0\degr-360\degr$). This gives 4 independent measurements of the normalized Stokes parameters $(q, u)$ in the $B$, $V$ and $R$ - bands. Such algorithm provides the best accuracy and helps to eliminate effects arising from e.g. dust particles on the retarder, non-parallelism of rotating components, etc.

The rotation speed of the modulator limits the maximum time resolution of the polarimeter.
\DUF\ performs one turn of the modulator by 22\fdg5 step in $\sim 0.25$~s, which corresponds to maximum rate of linear polarization measurement of $\sim$1.5~s$^{-1}$, provided that the signal from the target is high enough to allow exposure time $\ll 1$~s. For sufficiently bright stars this can be achieved even with the conventional amplifier. For fainter targets, EM mode can be employed in order to reduce exposure time while keeping S/N at the acceptable level. 

High sensitivity of the iXon EM CCD cameras and negligible light losses in the \DUF\ optics allow us to measure effectively polarization of faintest sources, even when mounted on a moderate-size telescope. This has proved to be useful in studying various transient sources, including low-mass X-ray binaries, which can be faint in the quiescent state. 

Each target is usually observed for a number of cycles, depending on brightness of the source, desired precision, and time resolution. The data is stored in the form of single-block single-channel FITS image files \citep{Wells1981}, producing independent data sets per each camera.
An example view of the sky is shown in Figure~\ref{fig:dipol_maxi_field}, left panel.
Each star within the field produces two orthogonally polarized images which are simultaneously recorded by the camera.

\subsection{Photometric mode}

\DUF\ can also operate in a photometric mode.
The analyzer (calcite) is removed from the optical path with the second stepper motor.  
The change of mode takes approximately 15 seconds, after which the instrument can be used as a three-band imaging photometer.
An image of the same region of the sky recorded in polarimetric and photometric modes can be seen in Figure~\ref{fig:dipol_maxi_field}.

In photometric mode the position angle of the modulator does not change, and there is no limit on the time resolution set by the the modulator's stepper motor.
Therefore, photometric regime allows full utilization of detectors' capabilities, including fast measurements of up to 56 full frames per second.

Apart from photometric measurements, a temporary retraction of calcite from the beam can be useful for target identification in the crowded fields. As is seen from Figure~\ref{fig:dipol_maxi_field}, the field around MAXI~J1820$+$070 can be difficult to navigate in polarimetric mode, and photometric mode can be used during the initial target acquisition. Photometric mode also reduces the minimum exposure time needed to identify a faint source, as the total incident flux is no longer split into two polarized beams, further simplifying target identification.

In crowded fields the polarized components may overlap with images of nearby field stars. This can be avoided by changing the field rotator angle to move the adjacent images apart.

\section{Control Hardware}

\begin{figure}
    \includegraphics[keepaspectratio, width = 1\linewidth]{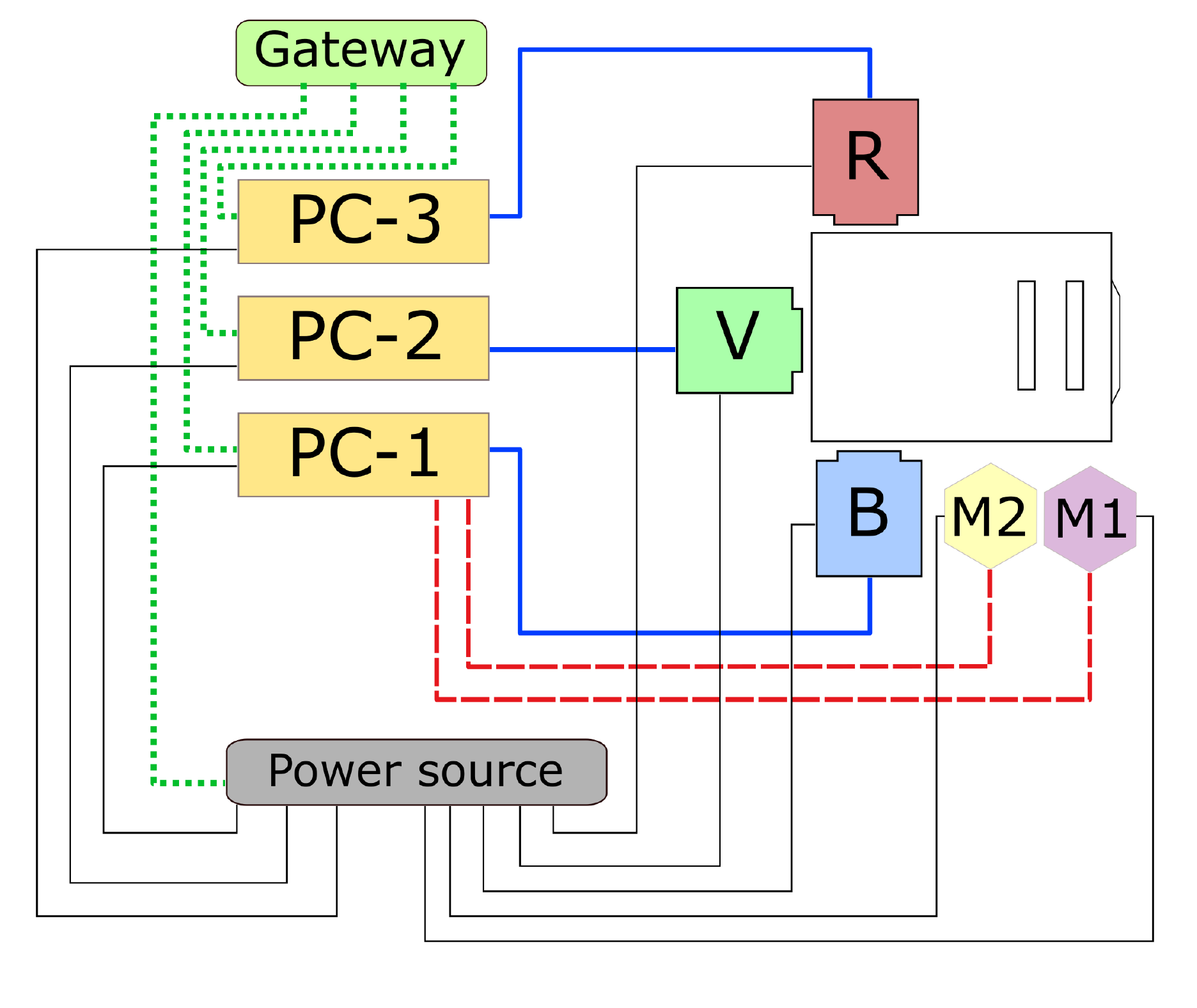}
    \caption{Schematic representation of \DUF\ hardware and their interconnections.
    `PC-1', `PC-2', and `PC-3' are control computers (`PC-1' is the main), `M1' is the first stepper motor, which rotates the modulator, `M2' is the second stepper motor, which controls position of the calcite plate.
    `B', `V', and `R' mark cameras for the respective passbands, `Power source' denotes the remotely-controlled power distribution unit (PDU), `Gateway' marks the network device (a router).
    Black solid lines show standard power cables, solid blue lines - universal serial bus (USB2) connections between computers and cameras, red dashed lines mark serial \citep[communication port, COM;][]{SPC2007} links to the stepper motors, green dotted lines show standard Ethernet cables \citep{Metcalfe1976}. }
    \label{fig:graph_hard}
\end{figure}

\DUF\ control hardware is based on three industrial-grade fanless computers to operate each of the three CCD cameras. We use Axiomtek eBox640-500-FL model\footnote{\url{https://www.axiomtek.com/Default.aspx?MenuId=Products&FunctionId=ProductView&ItemId=23389&C=eBOX640-500-FL&upcat=144}} as a primary (main) computer, which provides additional permanent storage for the acquired data, and control of two Trinamic stepper motors PD3-110-42.
Two identical Axiomtek eBox560-500-FL\footnote{\url{https://www.axiomtek.com/Default.aspx?MenuId=Products&FunctionId=ProductView&ItemId=17883&C=eBOX560-500-FL&upcat=144}} computers are used as secondary control PCs.
All computers are connected to a local area network, which is operated by a Mikrotik (Routerboard 2011UiAS-IN\footnote{\url{https://mikrotik.com/product/RB2011UiAS-IN}}) router.
We also use a remotely-controlled power distribution unit (PDU; Aviosys IP9258\footnote{\url{https://www.aviosys.com/products/9258.html}}), which allows us to remotely shutdown individual components of \DUF\ (such as cameras or computers). Hardware configuration is shown in Figure~\ref{fig:graph_hard}.

\DUF\ was built to be easily transported and mounted onto different telescopes. 
The router provides sufficient security between the external network and \DUF\ internal components. Each observer can reach \DUF\ computers by establishing a secure virtual private network (VPN) connection, which facilitates the use of the instrument by several observers from different locations. Our observing runs at the NOT in 2020 April and July were carried out in such a remote mode. Encrypted VPN connections protect both the instrument and the observer during an observing run, as well as the data transfer process.
 
\section{Control software}

\DUF\ consists of many elements that interact with each other.
In order to successfully operate the instrument we need an adequate control software.
We investigated existing third-party solutions, such as MaxIm DL,\footnote{\url{https://diffractionlimited.com/product/maxim-dl/}} which has been in use for several years with the \DP\ polarimeters, Andor Solis,\footnote{\url{https://andor.oxinst.com/products/solis-software/}} distributed with the Andor CCDs, and Astro.CONTROL.\footnote{\url{https://astrocontrol.de/en/astro-control-remote-camera-server/}}
After carefully assessing their performance in laboratory environment, we found that they, while providing an exceptional set of tools, are not optimal for controlling \DUF.
Both MaxIm DL and Andor Solis provide in-application scripting languages, but they are rather limited in options and do not support advanced scenarios, such as simultaneous and synchronous operation of three CCD cameras.

\begin{figure}
    \includegraphics[keepaspectratio, width = 1\linewidth]{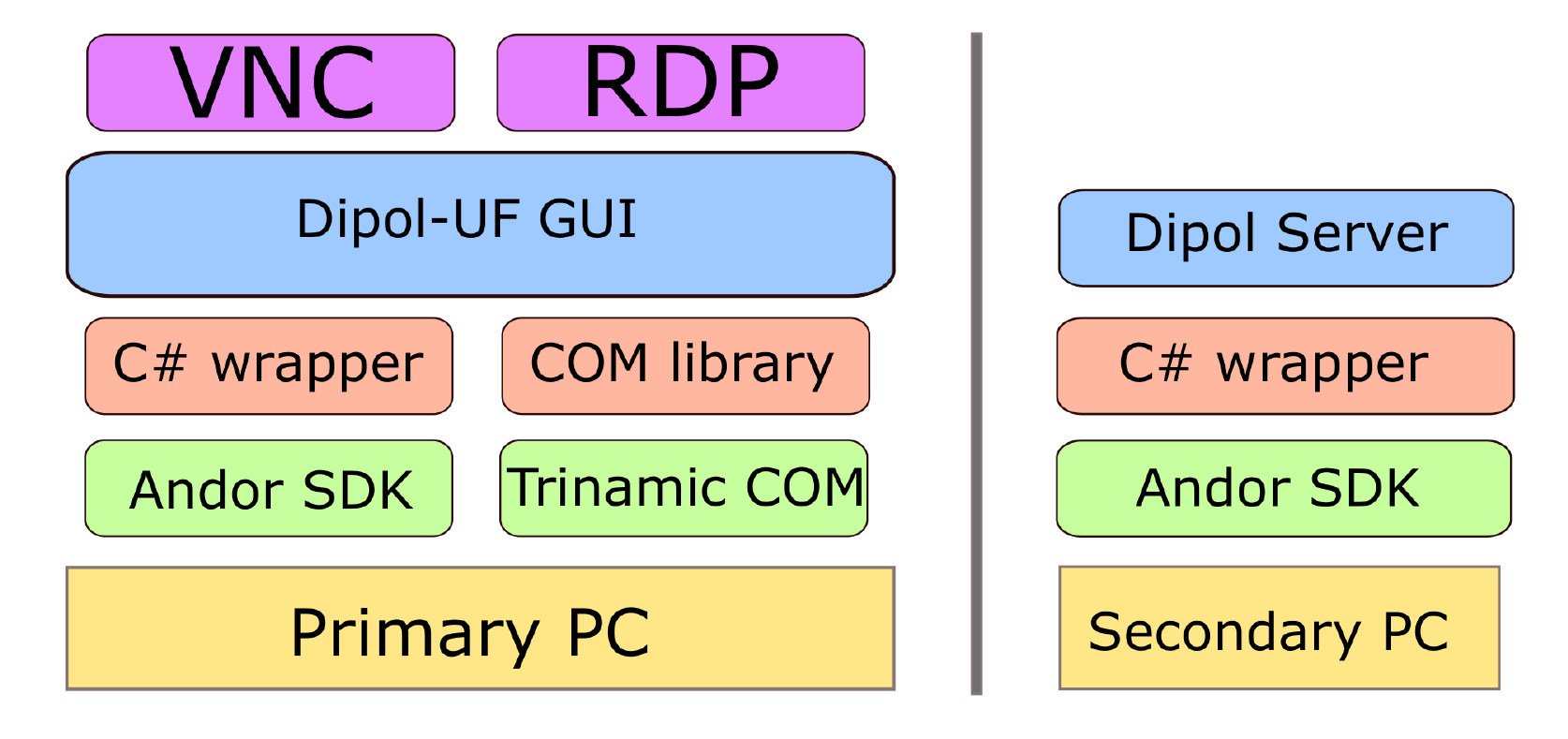}
    \caption{Layers of abstraction of the software, operating on the primary computer (`PC-1') and on the secondary computers (`PC-2' and `PC-3').
    At the bottom level orange box represents a 64-bit Windows NT 6.1-compatible operating system (OS; Windows 7 or newer) running on each computer.
    The green boxes are the proprietary Andor SDK, which provides access to CCD controls, and simple binary COM interface to the Trinamic stepper motors.
    The red boxes correspond to \DUF-specific higher-level libraries for interfacing with cameras and stepper motors, written in \code{\CS}.
    The blue boxes show the \DUF\ graphical user interface (GUI), which operates on the main computer, and special WCF services that run on the secondary computers, providing access to cameras over the network.
    Finally, the remote access to the primary computer is established using software such as Windows built-in remote desktop protocol (RDP) or third-party virtual network computing (VNC).
    }
    \label{fig:graph_soft}
\end{figure}

Preliminary analysis showed that a custom solution, built specifically for \DUF\ is required.
Fortunately, the manufacturer of the CCDs provides a rich application programming interface (API) via its software development kit\footnote{\url{https://andor.oxinst.com/products/software-development-kit/}} (SDK).
With the help of the SDK we were able to build an application suitable for our needs (see Figure~\ref{fig:graph_soft}).

The API is a set of simple \code{C} functions, which can be called from a variety of languages.
The SDK includes bindings to a number of popular languages, including \code{\CS}.\footnote{The most recent language standard is \href{https://www.ecma-international.org/publications/standards/Ecma-334.htm}{ECMA$-$334}.}
Such binding opens access to developing rich applications using Microsoft's .NET Framework.\footnote{The most recent runtime standard is \href{https://www.ecma-international.org/publications/standards/Ecma-335.htm}{ECMA$-$335}; the development of \DUF\ started before stable version of .NET Core was released.}

At the lowest level we developed a \code{\CS} wrapper library that helped us to effectively manage camera resources.
Each camera is represented as an object with properties describing the status of a camera and methods that issue commands to it.
We provided event-based notification of such properties as sensor temperature and cooling phase, and asynchronous (i.e. non-blocking) calls to time consuming (input/output bound) operations, such as waiting for an image acquisition to finish.
Our library also supports context switches between different cameras, allowing up to 8 cameras to be connected to one computer simultaneously (limited by the driver).
However, operating more than one camera at a time introduces some non-negligible overhead.

One of the main ideas introduced in the \code{\CS} library is the ability to manipulate camera configurations (acquisition settings).
Andor CCDs offer a lot of settings that can be adjusted before each image acquisition. 
We developed a dedicated format for storing acquisition settings in both human and machine-readable form.
The cameras can be preconfigured for each observational target before observing. 
New settings can be read from disk (or obtained from user input) and immediately sent to the cameras, greatly speeding up the process of switching targets.

We applied the same technique to develop the interface to the Trinamic stepper motors.
A subset of commands for operating each of the motors was implemented using a binary regime of communication over the serial port.
As with the cameras, each stepper motor is represented by an object on which methods can be asynchronously invoked to avoid blocking during rotations of the stepper motor.
This library is invoked from the main computer to control the rotation of the modulator and the position of the analyzer.

The next level of abstraction uses Windows Communication Foundation\footnote{\url{https://docs.microsoft.com/en-us/dotnet/framework/wcf/}} (WCF) to establish network communication between the main computer and the secondary computers.
Each secondary computer runs a WCF service, exposing its camera to the clients connected over the network. 
The main computer establishes connections to secondary computers, which allows it to manipulate `foreign' cameras that are not directly connected to it. Each of the `foreign' cameras is represented as a thin proxy, which is indistinguishable from the physically connected camera. 
This makes `local' and `foreign' cameras interchangeable. 
Any combination of physically connected and remotely operated cameras is supported, as long as each secondary computer runs a dedicated WCF service.

\begin{figure*}
\centering
    \includegraphics[keepaspectratio, width = 0.9\linewidth]{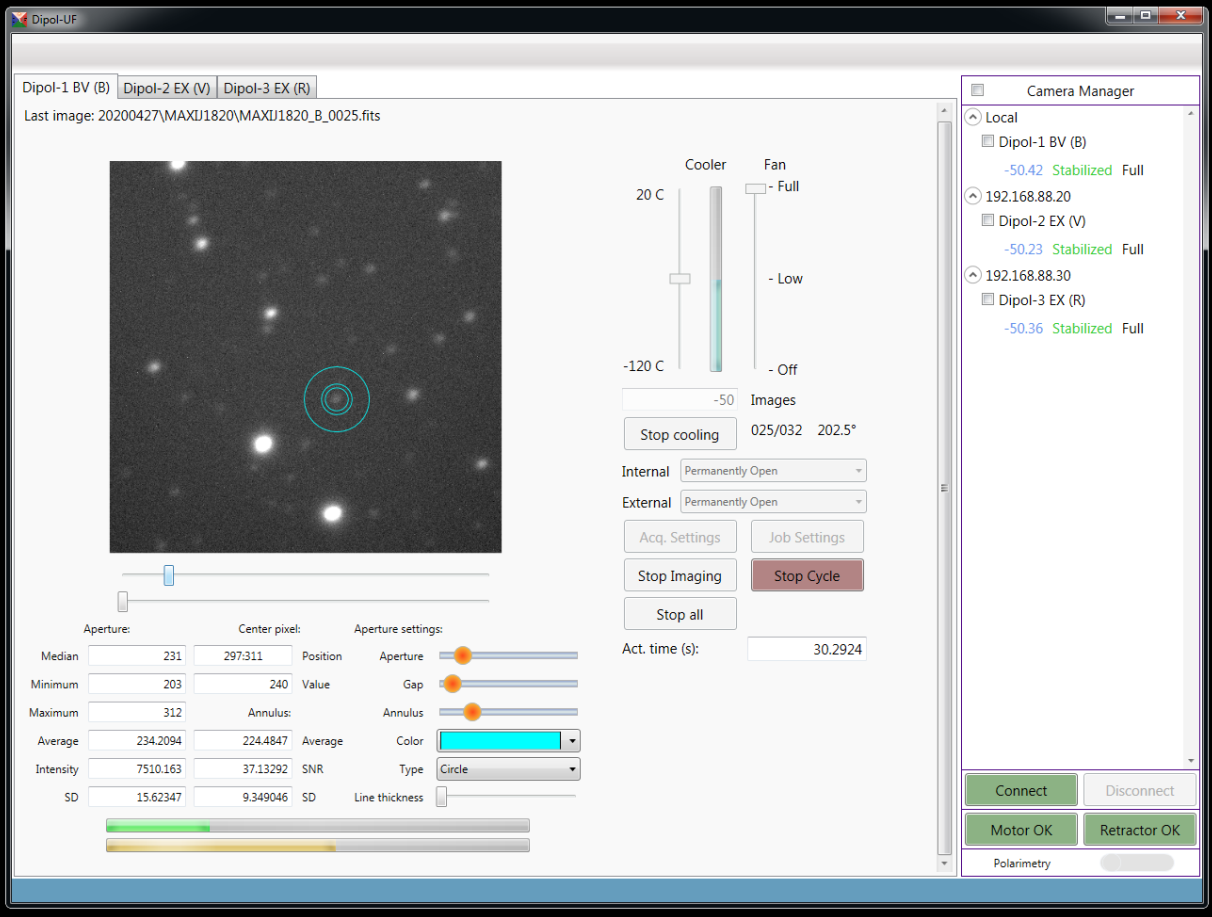}
    \caption{Main window of the \DUF\ GUI during polarimetric observations of MAXI~J1820$+$070.
    Rightmost panel shows connected Andor CCDs, their temperature and cooling phase.
    Central area shows a view on the $B$-filter camera controls, including the last obtained image (upper left corner), sampled image statistics (lower left corner), as well as acquisition and cycle controls (center screen).
    The progress bars at the bottom of the window show the progress of current exposure (green) and overall progress of the set of cycles (yellow).}
    \label{fig:gui}
\end{figure*}

Finally, we used Windows Presentation Foundation\footnote{\url{https://docs.microsoft.com/en-us/dotnet/framework/wpf/}} (WPF) to build a comprehensive graphical user interface (GUI), that provides observers with all necessary tools for controlling \DUF, including low-level camera controls (cooling, shutters, etc.), image acquisition and display, acquisition configuration (exposure time, vertical and horizontal shift speeds, amplifiers, acquisition, trigger and readout modes, etc.; both using the GUI and configuration files), and image conversion into FITS format. 
The \DUF\ GUI generates a separate tab for each camera, allowing the observer to quickly switch and inspect images obtained in different filters (Figure~\ref{fig:gui}).
Tabs also isolate camera configurations -- the majority of commands issued by the observer are received by the currently selected camera.
The synchronization between cameras is achieved through the `jobs' mechanism.
A job is a high-level configuration for \DUF\ -- it contains settings for each camera, information about instrument regime (polarimetric or photometric), and metadata about the target (such as target name and short textual description).

The job mechanism allows an observer to quickly switch between predefined configurations.
Every configuration loaded into \DUF\ is stored on the disk (JSON-compatible format) and can be reused later.
Alternatively, jobs can be pre-defined before the observations begin,  -- so that no time is wasted during the observation run.
We use a special job for target acquisition process -- it enables video mode on each camera, which turns single image display into continuous frame refresh, providing convenient way to adjust telescope focus and position the target in the image field.

\DUF\ control software is designed to prevent observer from accidental violation of the observation procedure.
The software takes care of the calibrations, automatically obtaining bias and dark images whenever the target and/or acquisition parameters changes.
This is important because changes in the CCD gain or horizontal/vertical shift speeds result in changes of CCD readout noise levels, brightness of the hot pixels, etc.

\DUF\ control software constantly monitors the state of each camera.
Each saved FITS image contains in its header detailed information about acquisition mode and acquisition parameters, camera temperature and angle of the modulator. This ensures that each image is correctly processed during data calibration and reduction phase and interrupted or unfinished cycles can be easily excluded from the final data set.

\section{Data processing and calibrations}

The images produced by \DUF\ require standard CCD calibrations \citep{Berdyugin2019}: bias and dark subtraction and flat field calibration.
The sky flats can be taken once per night.
Bias and dark exposure images are taken automatically at the end of the first set of cycles of a new target and saved with the target name prefixes, allowing their easy identification for the followup data calibration. If the acquisition parameters for the target were changed, the new set of calibration images is automatically taken.

During the first phase of data reduction, bias and dark images for the selected target are averaged (usually, a median image is computed) and then subtracted from the images of the target.
The flat-fielding can be also applied, however, its effect on the inferred value of polarization is minimal.

\begin{table*}[t]
\caption{
Normalized Stokes parameters $q$ and $u$ and their 1$\sigma$ errors for a sample of zero-polarization stars observed at the NOT. The data are given in units of $10^{-6}$ (ppm).
}
\label{tbl:low_stds}
\begin{tabular}{lccrrrcrrrcrrr}
        \hline
        \hline
        \multirow{2}{*}{Star} & \multirow{2}{*}{$m_V$\tablenotemark{a}} &
        \multirow{2}{*}{MJD} & \multicolumn{3}{c}{$B$} & & \multicolumn{3}{c}{$V$} & & \multicolumn{3}{c}{$R$} \\
\cline{4-6} \cline{8-10} \cline{12-14}
&          &  & $q$ & $u$ & $e$ & & $q$ & $u$ & $e$ & & $q$ & $u$ & $e$ \\ 
\hline
HD~4813   & 5.19 & 58687.200 &    3 &  $-$10 &   10 & &   $-$1 &   $-$2 &    9 & &   19 &   $-$2 &    9 \\
 HD~132052 & 4.49 & 58686.874 &  $-$24 &   21 &   14 & &   $-$1 &   $-$8 &   14 & &  12 &   23 &   13 \\
 HD~132254 & 5.60 & 58687.898 &    8 &  $-$12 &   13 & &   $-$3 &   $-$7 &   12 &  &  20 &   $-$5 &   10 \\
 HD~142373 & 4.62 & 58687.875 &    9 &   27 &   15 & &  19 &   15 &   10 & &    4 &   15 &   13 \\
 HD~147449 & 4.82 & 58687.917 &   $-$2 &   23 &   12 & &  $-$11 &    4 &   16  & &   28 &  $-$10 &   12 \\
 HD~187013 & 4.99 & 58688.078 &  $-$36 &  $-$10 &    12 & & $-$20 &  $-$23 &   12 & &   $-$9 &  $-$12 &   10 \\
 HD~187691 & 5.10 & 58687.040 &    1 &   $-$1 &   12 & &  $-$28 &  $-$41 &   11 & &  $-$15 &   $-$3 &    7 \\
 HD~191195 & 5.85 & 58687.064 &   $-$1 &   38 &   14 & &  $-$30 &   17 &   11 & &  $-$42 &   25 &   11 \\
 HD~205289 & 5.70 & 58687.083 &   $-$9 &   $-$9 &    13 & & $-$14 &   17 &   13 & &   12 &    4 &   12 \\
 HD~207958 & 5.08 & 58687.141 &  $-$25 &    3 &   12 & &  $-$12 &   $-$1 &   13 & &   17 &  $-$21 &   12 \\
 HD~213558 & 3.77 & 58688.201 &  $-$23 &   12 &   10 & &   17 &   $-$5 &    9 & &   14 &  $-$12 &   10 \\
 HD~215648 & 4.10 & 58687.118 &  $-$19 &  $-$31 &   12 & &  $-$20 &  $-$24 &    9 & &  $-$11 &  $-$12 &   10 \\
 HD~218470 & 5.70 & 58687.159 &  $-$10 &    7 &   12 & &   $-$7 &  $-$11 &    9 & &   17 &   $-$5 &    9 \\
 HD~219623 & 5.60 & 58688.221 &  $-$16 &   $-$9 &   12  & &   25 &  $-$14 &   11 & &   24 &    1 &   12 \\
 HD~225003 & 5.69 & 58687.178 &   $-$4 &  $-$29 &   10 &  &  $-$8 &  $-$17 &    8 & &   $-$3 &  $-$17 &    9 \\
 HD~189733\tablenotemark{b} & 7.65 & 58687.958 &   26 &   18 &    8 & &   10 &    0 &    7 & &   19 &   10 &    5 \\
 Telescope p.\tablenotemark{c} &   &  &    9 &  $-$48 &    3 & &   9 &  $-$83 &    3 &  &   0 &  $-$89 &    2 \\ 
    \hline
    \end{tabular}
\tablenotetext{a}{The $V$-band magnitudes are from SIMBAD database \citep{Wenger2000}.} 
\tablenotetext{b}{Star with a planetary component detected \citep{Bouchy2005, Berdyugina2011}.} 
\tablenotetext{c}{For alt-az telescope polarization produced by the telescope is given in telescope optics coordinates.}
\end{table*}

After calibrations, the difference in brightness of the ordinary and extraordinary rays of the target is extracted using differential aperture photometry method and Stokes parameters are computed from their intensity ratios \citep{Berdyugin2019}.
Individual values of Stokes parameters are then averaged using ``$2\sigma$'' weighting algorithm \citep[see][]{Kosenkov2017,Piirola2020}.
This method permits great flexibility in selecting how measurements are averaged, allowing observers to find balance between accuracy and time resolution.
For faint sources nightly average points are commonly used \citep[see examples from \DP\ in][]{Kosenkov2017, Veledina2019,Kosenkov2020b}.

If there are other stars in the field around the target within the CCD frame, their polarization can be also measured and used for the determination of
the interstellar (ISM) polarization in the direction of the target.
Field star polarization and parallaxes \citep[e.g., from \textit{Gaia} data release \#2;][]{Gaia2018} provide an excellent tool for studying dependence of the ISM polarization on distance. 
The value of the (ISM) polarization component can be subtracted from the observed polarization, to get the intrinsic polarization of the target -- a valuable source of information about the polarizing mechanism and the physical and geometrical properties of the object.

\section{First results}
\label{sec:first}

\begin{table*}
\caption{Sample of high-polarization stars observed at the NOT. }
    \label{tbl:high_stds}
    \centering
    \begin{tabular}{lcr@{$\,\pm\,$}lr@{$\,\pm\,$}lcr@{$\,\pm\,$}lr@{$\,\pm\,$}lcr@{$\,\pm\,$}lr@{$\,\pm\,$}l}
        \hline
        \hline
        \multirow{2}{*}{Star} & \multirow{2}{*}{Date} &       \multicolumn{4}{c}{$B$} & & \multicolumn{4}{c}{$V$} & & \multicolumn{4}{c}{$R$} \\
\cline{3-6}  \cline{8-11}  \cline{13-16}
                              &                        & \multicolumn{2}{c}{$P$} & \multicolumn{2}{c}{$\theta$} && \multicolumn{2}{c}{$P$} & \multicolumn{2}{c}{$\theta$} & & \multicolumn{2}{c}{$P$} & \multicolumn{2}{c}{$\theta$} \\
                              &     (MJD)           & \multicolumn{2}{c}{(\%)} & \multicolumn{2}{c}{(deg)} && \multicolumn{2}{c}{(\%)} & \multicolumn{2}{c}{(deg)}  && \multicolumn{2}{c}{(\%)} & \multicolumn{2}{c}{(deg)} \\
        \hline

        \multirow{3}{*}{HD~161056}  & 59051.887 & 3.769 & 0.004 &  67.06 &  0.03 && 4.051 & 0.003 &  67.15 &  0.02 & & 3.981 & 0.012 &  67.71 &  0.09 \\
        & 59052.888 & 3.760 & 0.004 &  67.15 &  0.03 && 4.044 & 0.004 &  67.24 &  0.03 & &  4.008  & 0.003 &  67.76 &  0.02 \\
        & 59053.886 & 3.763 & 0.004 &  67.15 &  0.03 && 4.042 & 0.004 &  67.15 &  0.03 & & 4.010 & 0.003 &  67.69 &  0.02 \\
   average  &   & 3.764 & 0.002 &  67.12 &  0.02 & & 4.046 & 0.002 &  67.18 &  0.01 &  & 4.009 & 0.002 &  67.72 &  0.01\\
        \multicolumn{16}{c}{ } \\ 
         \multirow{4}{*}{HD~204827}  & 59051.210 & 5.789 & 0.011 &  57.57 &  0.05 & & 5.602 & 0.019 &  58.08 &  0.10 & & 5.079 & 0.011 &  58.97 &  0.06 \\
  & 59052.211 & 5.787 & 0.004 &  57.67 &  0.02 & & 5.609 & 0.009 &  58.12 &  0.05 & & 5.084 & 0.005 &  59.09 &  0.03 \\
  & 59053.213 & 5.783 & 0.007 &  57.92 &  0.03 & & 5.608 & 0.004 &  58.35 &  0.02 & & 5.089 & 0.005 &  59.30 &  0.03 \\
  & 59054.213 & 5.788 & 0.006 &  58.02 &  0.03 & & 5.605 & 0.005 &  58.39 &  0.03 & & 5.086 & 0.004 &  59.29 &  0.02 \\
  average    &   & 5.786 & 0.004 &  57.79 &  0.02 & & 5.607 & 0.003 &  58.33 &  0.02 & & 5.085 & 0.003 &  59.21 &  0.02 \\
                                   
        \hline
    \end{tabular}\
\tablecomments{Errors are 1$\sigma$.}
\end{table*}

\begin{table*}
\caption{Polarization properties of a sample of science targets observed at the NOT.}
    \label{tbl:transients}
    \centering
    \begin{tabular}{ccccr@{$\,\pm\,$}lr@{$\,\pm\,$}lr@{$\,\pm\,$}lr@{$\,\pm\,$}lr@{$\,\pm\,$}lr@{$\,\pm\,$}l}
        \hline
        \hline
        \multirow{2}{*}{Star} & \multirow{2}{*}{$m_V$\tablenotemark{a}} & \multirow{2}{*}{Date} & Total & \multicolumn{4}{c}{$B$} & \multicolumn{4}{c}{$V$} & \multicolumn{4}{c}{$R$} \\
\cline{5-7}  \cline{9-11}  \cline{13-16}
                            &                       & & exp. & \multicolumn{2}{c}{$P$} & \multicolumn{2}{c}{$\theta$} & \multicolumn{2}{c}{$P$} & \multicolumn{2}{c}{$\theta$} & \multicolumn{2}{c}{$P$} & \multicolumn{2}{c}{$\theta$} \\
                            &         (mag)          & (MJD) & (min) & \multicolumn{2}{c}{(\%)} & \multicolumn{2}{c}{(deg)} & \multicolumn{2}{c}{(\%)} & \multicolumn{2}{c}{(deg)} & \multicolumn{2}{c}{(\%)} & \multicolumn{2}{c}{(deg)} \\
        \hline
        V404 Cyg\tablenotemark{b} & $>$18.0& \multirow{2}{*}{58688.019} & \multirow{2}{*}{{32}} & \multicolumn{2}{c}{...} & \multicolumn{2}{c}{...} & 8.07 & 0.41 & 3.1 & 1.5 & 7.30 & 0.14 & 5.3 & 0.5 \\
       Companion & &  & &\multicolumn{2}{c}{...} & \multicolumn{2}{c}{...} & 8.65 & 0.42 & $-$0.3 & 1.4 & 7.45 & 0.25 & 5.1 & 1.0 \\
        \multicolumn{15}{c}{ } \\
         \multirow{2}{*}{MAXI~J1820} & \multirow{2}{*}{18.5}  & 58686.927 & \multirow{2}{*}{{32}} & 3.05 & 0.29 & 156.7 & 2.7 & 1.10 & 0.24 & 154.8 & 6.2 & 0.11 & 0.14 & 164.1 & 26.1 \\
         & & 58688.014 & & 1.54 & 0.38 & 154.2 & 7.0 & 0.95 & 0.25 & 28.0 & 7.3 & 1.24 & 0.10 & 129.9 & \ 2.3 \\
            \multicolumn{15}{c}{ } \\  
        SS~433 & 14.5 & 58687.037 & {32} & 4.29 & 0.20 & $-$3.7 & 1.3 & 4.38 & 0.07 & $-$3.9 & 0.5 & 4.46 & 0.03 & $-$3.1 & 0.2 \\
        \multicolumn{15}{c}{ } \\ 
        \multirow{5}{*}{V4641~Sgr} & \multirow{5}{*}{13.5} 
  & 58686.984 & {10}  & 0.53 & 0.03 & 43.6 & 1.6 & 0.40 & 0.04 & 46.0 & 2.7 & 0.45 & 0.02 & 50.7 & 0.9 \\
& & 58961.189 & {32} & 0.46 & 0.02 & 36.8 & 1.4 & 0.49 & 0.06 & 45.7 & 3.6 & 0.42 & 0.06 & 48.7 & 4.3 \\
& & 58964.211 & {64} & 0.49 & 0.01 & 40.2 & 0.8 & 0.43 & 0.03 & 44.7 & 1.8 & 0.44 & 0.02 & 50.7 & 1.0 \\
& & 58966.215 & {40} & 0.50 & 0.03 & 40.1 & 1.4 & 0.44 & 0.04 & 46.0 & 2.3 & 0.48 & 0.04 & 50.0 & 2.4 \\
& & 58967.200 & {26} & 0.51 & 0.05 & 44.4 & 2.9 & 0.50 & 0.10 & 43.2 & 5.7 & 0.45 & 0.10 & 51.2 & 6.1 \\
        Field star\tablenotemark{c} & & 58964.088 & {162} & 0.47 & 0.02 & 41.0 & 1.1 &\multicolumn{2}{c}{...} & \multicolumn{2}{c}{...} &\multicolumn{2}{c}{...} & \multicolumn{2}{c}{...}\\
        \multicolumn{15}{c}{ } \\
        \multirow{4}{*}{WD~1145$+$017} & \multirow{4}{*}{17.0} & 58961.037 & {128}  & 0.20 & 0.06 & 29.9 & \ph{1}7.6 & 0.04 & 0.07 & 43.6 & 30.7 & 0.18 & 0.07 & 38.9 & 10.4\\
         & & 58964.052 & {245} & 0.14 & 0.04 & 9.4 & \ph{1}7.9 & 0.18 & 0.05 & 169.9 & \ph{1}8.5 & 0.14 & 0.06 & 13.4 & 11.1 \\
         & & 58964.967 & {256} & 0.10 & 0.04 & 27.7 & 10.3 & 0.10 & 0.05 & 10.0  & 12.0 & 0.12 & 0.05 & 5.1 & 11.6 \\
         & & 58965.976 & {256} & 0.10 & 0.04 & 175.4 & 11.7 & 0.09 & 0.05 & 9.5 & 13.3 & 0.11 & 0.05 & 22.4 & 12.7 \\
        \hline
    \end{tabular}
    \tablecomments{Errors are 1$\sigma$.}
    \tablenotetext{a}{We use publicly available AAVSO data \citep{Kafka2020} to estimate $V$-band magnitude of the stars during our observing runs.}
    \tablenotetext{b}{$B$ filter data for V404~Cyg are lacking due to large interstellar extinction \citep{Shahbaz2003} and nearby close companion \citep{Casares2014}.}
    \tablenotetext{c}{Both images (o and e) of the brightest nearby field star are only visible in the $B$ band frames.}
\end{table*}

\DUF\ has been commissioned at the Nordic Optical Telescope (NOT) in July 22--23, 2019 and used for regular observing runs in April 21--27 and July 20--24, 2020.  
Among the main targets were the low-mass X-ray binaries V404~Cyg \citep{Khargharia2010}  and MAXI~J1820$+$070 \citep{Torres2020}, an intermediate-mass X-ray binary  V4641~Sgr \citep{Orosz2001,MacDonald2014}, a high-mass X-ray binary SS~433 \citep{Fabrika2004} and white dwarf WD~1145$+$017 with a debris disk \citep{Vanderburg2015}.

In each observing run we measured also a sample of low-polarization nearby standard stars to determine instrumental polarization (see Table~\ref{tbl:low_stds}). Polarization produced by the optics of the NOT telescope is low ($< 10^{-4}$) and stable, and was subtracted from the science target observations. No significant instrumental polarization has been found from the polarimeter itself. The experience from the first observations is that precision at $10^{-5}$ or better for bright stars is achieved in considerably shorter time (30--40 min) than with the conventional CCD camera polarimeter \DP\ ($\sim$ 1 hour).

Broad-band polarization data derived by averaging the 
normalized Stokes parameters obtained in the $B$, $V$, and $R$ bands  
(Table~\ref{tbl:low_stds}) have been used by \citet{Piirola2020}  
for mapping interstellar dust and magnetic field inside the Local Bubble.
High-precision polarimetric observations of nearby stars (at the ppm level)
have recently been published also by \citet{Bailey10} and \citet{Cotton17, Cotton19}. 

We have also observed two high-polarization standard stars HD~161056 and HD~204827 (Table~\ref{tbl:high_stds}) to calibrate the zero point of polarization angle (PA). 
From the measurements shown for four nights in July 20--24, 2020, consistent values for both the degree of polarization (PD) and PA were obtained. 
The time spent on each nightly observation was only 5--10 min, since the highest S/N was not needed for the large polarization standards. The PA values agree within about 0\fdg5 with those given in the large polarization standard star list by \citet{Schmidt92}. Also the PD for HD~161056 is in fair agreement with these published data. For HD~204827 our PD values are larger by $\sim$ 0.2\% than in \citet{Schmidt92}, and larger by $\sim$ 0.1\% than in \citet{Hsu82}, suggesting possible contribution from stellar intrinsic polarization variations. The difference cannot be due to instrumental effects in \DUF\ since any deviation of the retardance of the HWP from $\lambda/2$ will {\it reduce} the measured polarization value. Evidence for polarization variability in HD~204827 has been found already by \citet{DolanTapia86}, and polarization changes up to $\Delta P \sim 0.226\%, \Delta \theta \sim 0\fdg6$, have been reported by \citet{Bastien88}.

Polarization properties of a sample of science targets observed at the NOT are given in Table~\ref{tbl:transients}. The total exposure times are shown in the fourth column. Telescope pointing, target acquisition, and dark exposures add several minutes to the total telescope time. With individual exposure times 10--30~s for faint targets, the time lost between exposures ($\sim$0.3~s) is negligible.

A black hole X-ray binary V404~Cyg was extensively monitored during its June 2015 outburst \citep[see, e.g.,][]{Tanaka2016, Shahbaz2016, Itoh2017, Kosenkov2017}, and variable intrinsic polarization was reported. 
The intrinsic polarization was calculated by subtracting the average value of V404~Cyg polarization right after the outburst has ended \citep{Kosenkov2017}, assuming that polarization is purely of interstellar origin. 
During our first commissioning run of \DUF\ in July 2019, we briefly observed V404~Cyg in its quiescent state.
Our observations of V404~Cyg and its close companion star (1\farcs4 away) show very similar polarization, which strongly suggests that the quiescent state polarization is of interstellar origin. This is an important result for the interpretation of the variable polarization observed  during the outburst of V404~Cyg. 

We monitored MAXI~J1820$+$070 in polarized light during its 2018 outburst with \DP\ \citep{Veledina2019, Kosenkov2020b} and detected a weak intrinsic polarization below 0.5\%. 
In July 2019 we observed the object in its quiescent state with \DUF. 
In contrast to V404~Cyg, and for the first time, we detected a significant intrinsic polarization from a black hole X-ray binary in quiescence, reaching 3\% in the $B$ band and decreasing toward the $R$ band (see Table~\ref{tbl:transients}). 
The steep wavelength dependence of polarization with the peak in the blue implies a thermal polarized source. 
 
The polarization of SS~433 measured in July 2019 was relatively high, but shows no peculiarities. 
Both the polarization degree and angle agree within errors in all three bands, suggesting that polarization is produced mostly by the ISM. 
This is in agreement with the value of interstellar polarization derived by \citet{Efimov1984} from the high-state observations of SS~433, and indicates that the object was in a low state during our observations.

V4641~Sgr is a unique intermediate-mass X-ray binary system hosting a $\sim6.4M_\odot$ black hole and a $\sim 2.9M_\odot$ early type (B9III) donor star in a 2.82~d orbit  \citep{MacDonald2014}.
The relatively short orbital period makes V4641~Sgr a convenient target for our polarimetric campaign. 
Our \DUF\ observations of V4641~Sgr (see Table~\ref{tbl:transients}) show that polarization of the binary system remains stable on different time-scales: first measurement was obtained almost a year before the last four observations, which covered two orbital revolutions.
From the same exposures we were also able to measure polarization of a bright field star in the $B$-filter.
Both the polarization degree and polarization angle of the field star agree with those of V4641~Sgr, suggesting that the binary system exhibits little or no intrinsic polarization in the quiescent state.

WD 1145+017 is a white dwarf with a debris disk and quasi-periodic brightness dips which most likely occur due to dust cloud transits \citep{Vanderburg2018}. We measured the polarization of this target to probe the physical properties of the disk material. 
First measurements made with \DUF\ at the NOT (Table~\ref{tbl:transients}) show small, but non-zero polarization at the level of $\sim0.15 \%$ with possible night-to-night variability.

\section{Conclusions}

The new version of our double-image polarimeters, \DUF, has proven the power of fast-readout EM CCDs for high-precision polarimetric observations. 
In this paper we demonstrated the capabilities of the instrument showing some examples covering a wide range of applications, from the studies of the minute polarization produced by the interstellar dust and magnetic field in the Solar neighborhood, to black hole X-ray binaries. 
Other potential targets are rapidly variable objects, such as Cataclysmic Variables with extremely strong magnetic fields (mCVs). 
For these objects simultaneous measurements in multiple wavebands are essential. With the precision of $10^{-5}$ and better, tighter constraints can be put also on the polarized flux reflected from exoplanet atmospheres.
\DUF\ is available for polarimetric research for a wide international community at the NOT telescope.

\acknowledgements
\DUF\ is a joint effort between University of Turku (Finland) and Leibniz Institute for Solar Physics (Germany). 
We acknowledge support from the Magnus Ehrnrooth foundation and ERC Advanced Grant HotMol ERC-2011-AdG-291659. 
Based on observations made with the Nordic Optical Telescope, owned in collaboration by the University of Turku and Aarhus University, and operated jointly by Aarhus University, the University of Turku and the University of Oslo, representing Denmark, Finland and Norway, the University of Iceland and Stockholm University at the Observatorio del Roque de los Muchachos, La Palma, Spain, of the Instituto de Astrofisica de Canarias.
We thank the NOT staff for the excellent support.
\vspace{5mm}

\facility{NOT}
\vspace{62mm}
\bibliography{references}{}

\begin{thebibliography}{}
\expandafter\ifx\csname natexlab\endcsname\relax\def\natexlab#1{#1}\fi
\providecommand{\url}[1]{\href{#1}{#1}}
\providecommand{\dodoi}[1]{doi:~\href{http://doi.org/#1}{\nolinkurl{#1}}}
\providecommand{\doeprint}[1]{\href{http://ascl.net/#1}{\nolinkurl{http://ascl.net/#1}}}
\providecommand{\doarXiv}[1]{\href{https://arxiv.org/abs/#1}{\nolinkurl{https://arxiv.org/abs/#1}}}

\bibitem[{{Arnold} {et~al.}(2012){Arnold}, {Steele}, {Bates}, {Mottram}, \&
  {Smith}}]{Arnold2012}
{Arnold}, D.~M., {Steele}, I.~A., {Bates}, S.~D., {Mottram}, C.~J., \& {Smith},
  R.~J. 2012, in \procspie, Vol. 8446, Ground-based and Airborne
  Instrumentation for Astronomy IV, 84462J, \dodoi{10.1117/12.927000}

\bibitem[{{Axelson}(2007)}]{SPC2007}
{Axelson}, J. 2007, Serial Port Complete (Madison: Lakeview Research)

\bibitem[{{Bailey} {et~al.}(2010){Bailey}, {Lucas}, \& {Hough}}]{Bailey10}
{Bailey}, J., {Lucas}, P.~W., \& {Hough}, J.~H. 2010, \mnras, 405, 2570,
  \dodoi{10.1111/j.1365-2966.2010.16634.x}

\bibitem[{{Bastien} {et~al.}(1988){Bastien}, {Drissen}, {Menard}, {Moffat},
  {Robert}, \& {St-Louis}}]{Bastien88}
{Bastien}, P., {Drissen}, L., {Menard}, F., {et~al.} 1988, \aj, 95, 900,
  \dodoi{10.1086/114688}

\bibitem[{{Berdyugin} {et~al.}(2019){Berdyugin}, {Piirola}, \&
  {Poutanen}}]{Berdyugin2019}
{Berdyugin}, A., {Piirola}, V., \& {Poutanen}, J. 2019, in Astrophysics and
  Space Science Library, Vol. 460, Astronomical Polarisation from the Infrared
  to Gamma Rays, ed. R.~{Mignani}, A.~{Shearer}, A.~{S{\l}owikowska}, \&
  S.~{Zane} (Cham: Springer Nature), 33--65,
  \dodoi{10.1007/978-3-030-19715-5_3}

\bibitem[{{Berdyugina} {et~al.}(2011){Berdyugina}, {Berdyugin}, {Fluri}, \&
  {Piirola}}]{Berdyugina2011}
{Berdyugina}, S.~V., {Berdyugin}, A.~V., {Fluri}, D.~M., \& {Piirola}, V. 2011,
  \apjl, 728, L6, \dodoi{10.1088/2041-8205/728/1/L6}

\bibitem[{{Bouchy} {et~al.}(2005){Bouchy}, {Udry}, {Mayor}, {Moutou}, {Pont},
  {Iribarne}, {da Silva}, {Ilovaisky}, {Queloz}, {Santos}, {S{\'e}gransan}, \&
  {Zucker}}]{Bouchy2005}
{Bouchy}, F., {Udry}, S., {Mayor}, M., {et~al.} 2005, \aap, 444, L15,
  \dodoi{10.1051/0004-6361:200500201}

\bibitem[{{Casares} \& {Jonker}(2014)}]{Casares2014}
{Casares}, J., \& {Jonker}, P.~G. 2014, \ssr, 183, 223,
  \dodoi{10.1007/s11214-013-0030-6}

\bibitem[{{Clarke}(1965)}]{Clarke1965}
{Clarke}, D. 1965, \mnras, 130, 75, \dodoi{10.1093/mnras/130.1.75}

\bibitem[{{Collins} {et~al.}(2013){Collins}, {Kyne}, {Lara}, {Redfern},
  {Shearer}, \& {Sheehan}}]{Collin2013}
{Collins}, P., {Kyne}, G., {Lara}, D., {et~al.} 2013, Experimental Astronomy,
  36, 479, \dodoi{10.1007/s10686-013-9342-5}

\bibitem[{{Cotton} {et~al.}(2017){Cotton}, {Marshall}, {Bailey},
  {Kedziora-Chudczer}, {Bott}, {Marsden}, \& {Carter}}]{Cotton17}
{Cotton}, D.~V., {Marshall}, J.~P., {Bailey}, J., {et~al.} 2017, \mnras, 467,
  873, \dodoi{10.1093/mnras/stx068}

\bibitem[{{Cotton} {et~al.}(2019){Cotton}, {Marshall}, {Frisch},
  {Kedziora-Chudzer}, {Bailey}, {Bott}, {Wright}, {Wyatt}, \&
  {Kennedy}}]{Cotton19}
{Cotton}, D.~V., {Marshall}, J.~P., {Frisch}, P.~C., {et~al.} 2019, \mnras,
  483, 3636, \dodoi{10.1093/mnras/sty3318}

\bibitem[{{Denisenko}(2018)}]{Denisenko2018}
{Denisenko}, D. 2018, The Astronomer's Telegram, 11400, 1

\bibitem[{{Dolan} \& {Tapia}(1986)}]{DolanTapia86}
{Dolan}, J.~F., \& {Tapia}, S. 1986, \pasp, 98, 792, \dodoi{10.1086/131827}

\bibitem[{{Efimov} {et~al.}(1984){Efimov}, {Shakhovskoi}, \&
  {Piirola}}]{Efimov1984}
{Efimov}, I.~S., {Shakhovskoi}, N.~M., \& {Piirola}, V. 1984, \aap, 138, 62

\bibitem[{{Fabrika}(2004)}]{Fabrika2004}
{Fabrika}, S. 2004, \apspr, 12, 1.
\newblock \doarXiv{astro-ph/0603390}

\bibitem[{{Gaia Collaboration} {et~al.}(2018)}]{Gaia2018}
{Gaia Collaboration}, {et~al.} 2018, \aap, 616, A1,
  \dodoi{10.1051/0004-6361/201833051}

\bibitem[{{Gisler} {et~al.}(2016){Gisler}, {Berkefeld}, \&
  {Berdyugina}}]{Gisler2016}
{Gisler}, D., {Berkefeld}, T., \& {Berdyugina}, S. 2016, in \procspie, Vol.
  9906, Ground-based and Airborne Telescopes VI, 99065E,
  \dodoi{10.1117/12.2233461}

\bibitem[{{Hsu} \& {Breger}(1982)}]{Hsu82}
{Hsu}, J.~C., \& {Breger}, M. 1982, \apj, 262, 732, \dodoi{10.1086/160467}

\bibitem[{{Itoh} {et~al.}(2017){Itoh}, {Tanaka}, {Kawabata}, {Uemura},
  {Watanabe}, {Fukazawa}, {Kand a}, {Akitaya}, {Moritani}, {Nakaoka},
  {Kawabata}, {Shiki}, {Yoshida}, {Oasa}, \& {Takahashi}}]{Itoh2017}
{Itoh}, R., {Tanaka}, Y.~T., {Kawabata}, K.~S., {et~al.} 2017, \pasj, 69, 25,
  \dodoi{10.1093/pasj/psw130}

\bibitem[{{Kafka}(2020)}]{Kafka2020}
{Kafka}, S. 2020, Observations from the AAVSO International Database.
\newblock \url{https://www.aavso.org/}

\bibitem[{{Kawamuro} {et~al.}(2018){Kawamuro}, {Negoro}, {Yoneyama}, {Ueno},
  {Tomida}, {Ishikawa}, {Sugawara}, {Isobe}, {Shimomukai}, {Mihara},
  {Sugizaki}, {Nakahira}, {Iwakiri}, {Yatabe}, {Takao}, {Matsuoka}, {Kawai},
  {Sugita}, {Yoshii}, {Tachibana}, {Harita}, {Morita}, {Yoshida}, {Sakamoto},
  {Serino}, {Kawakubo}, {Kitaoka}, {Hashimoto}, {Tsunemi}, {Nakajima},
  {Kawase}, {Sakamaki}, {Maruyama}, {Ueda}, {Hori}, {Tanimoto}, {Oda},
  {Morita}, {Yamada}, {Tsuboi}, {Nakamura}, {Sasaki}, {Kawai}, {Sato},
  {Yamauchi}, {Hanyu}, {Hidaka}, {Yamaoka}, \& {Shidatsu}}]{Kawamuro2018}
{Kawamuro}, T., {Negoro}, H., {Yoneyama}, T., {et~al.} 2018, The Astronomer's
  Telegram, 11399, 1

\bibitem[{{Khargharia} {et~al.}(2010){Khargharia}, {Froning}, \&
  {Robinson}}]{Khargharia2010}
{Khargharia}, J., {Froning}, C.~S., \& {Robinson}, E.~L. 2010, \apj, 716, 1105,
  \dodoi{10.1088/0004-637X/716/2/1105}

\bibitem[{{Kosenkov} {et~al.}(2017){Kosenkov}, {Berdyugin}, {Piirola},
  {Tsygankov}, {Pall{\'e}}, {Miles-P{\'a}ez}, \& {Poutanen}}]{Kosenkov2017}
{Kosenkov}, I.~A., {Berdyugin}, A.~V., {Piirola}, V., {et~al.} 2017, \mnras,
  468, 4362, \dodoi{10.1093/mnras/stx779}

\bibitem[{{Kosenkov} {et~al.}(2020){Kosenkov}, {Veledina}, {Berdyugin},
  {Kravtsov}, {Piirola}, {Berdyugina}, {Sakanoi}, {Kagitani}, \&
  {Poutanen}}]{Kosenkov2020b}
{Kosenkov}, I.~A., {Veledina}, A., {Berdyugin}, A.~V., {et~al.} 2020, \mnras,
  496, L96, \dodoi{10.1093/mnrasl/slaa096}

\bibitem[{{MacDonald} {et~al.}(2014){MacDonald}, {Bailyn}, {Buxton},
  {Cantrell}, {Chatterjee}, {Kennedy-Shaffer}, {Orosz}, {Markwardt}, \&
  {Swank}}]{MacDonald2014}
{MacDonald}, R. K.~D., {Bailyn}, C.~D., {Buxton}, M., {et~al.} 2014, \apj, 784,
  2, \dodoi{10.1088/0004-637X/784/1/2}

\bibitem[{{Metcalfe} \& {Boggs}(1976)}]{Metcalfe1976}
{Metcalfe}, R.~M., \& {Boggs}, D.~R. 1976, Commun. ACM, 19, 395–404,
  \dodoi{10.1145/360248.360253}

\bibitem[{{Orosz} {et~al.}(2001){Orosz}, {Kuulkers}, {van der Klis},
  {McClintock}, {Garcia}, {Callanan}, {Bailyn}, {Jain}, \&
  {Remillard}}]{Orosz2001}
{Orosz}, J.~A., {Kuulkers}, E., {van der Klis}, M., {et~al.} 2001, \apj, 555,
  489, \dodoi{10.1086/321442}

\bibitem[{{Piirola}(1973)}]{Piirola1973}
{Piirola}, V. 1973, \aap, 27, 383

\bibitem[{{Piirola} {et~al.}(2014){Piirola}, {Berdyugin}, \&
  {Berdyugina}}]{Piirola2014}
{Piirola}, V., {Berdyugin}, A., \& {Berdyugina}, S. 2014, in \procspie, Vol.
  9147, Ground-based and Airborne Instrumentation for Astronomy V, 91478I,
  \dodoi{10.1117/12.2055923}

\bibitem[{{Piirola} {et~al.}(2020){Piirola}, {Berdyugin}, {Frisch}, {Kagitani},
  {Sakanoi}, {Berdyugina}, {Cole}, {Harlingten}, \& {Hill}}]{Piirola2020}
{Piirola}, V., {Berdyugin}, A., {Frisch}, P.~C., {et~al.} 2020, \aap, 635, A46,
  \dodoi{10.1051/0004-6361/201937324}

\bibitem[{{Schmidt} {et~al.}(1992){Schmidt}, {Elston}, \& {Lupie}}]{Schmidt92}
{Schmidt}, G.~D., {Elston}, R., \& {Lupie}, O.~L. 1992, \aj, 104, 1563,
  \dodoi{10.1086/116341}

\bibitem[{{Shahbaz} {et~al.}(2003){Shahbaz}, {Dhillon}, {Marsh}, {Zurita},
  {Haswell}, {Charles}, {Hynes}, \& {Casares}}]{Shahbaz2003}
{Shahbaz}, T., {Dhillon}, V.~S., {Marsh}, T.~R., {et~al.} 2003, \mnras, 346,
  1116, \dodoi{10.1111/j.1365-2966.2003.07158.x}

\bibitem[{{Shahbaz} {et~al.}(2016){Shahbaz}, {Russell}, {Covino}, {Mooley},
  {Fender}, \& {Rumsey}}]{Shahbaz2016}
{Shahbaz}, T., {Russell}, D.~M., {Covino}, S., {et~al.} 2016, \mnras, 463,
  1822, \dodoi{10.1093/mnras/stw2171}

\bibitem[{{Tanaka} {et~al.}(2016){Tanaka}, {Itoh}, {Uemura}, {Inoue}, {Cheung},
  {Watanabe}, {Kawabata}, {Fukazawa}, {Yatsu}, {Yoshii}, {Tachibana},
  {Fujiwara}, {Saito}, {Kawai}, {Kimura}, {Isogai}, {Kato}, {Akitaya},
  {Kawabata}, {Nakaoka}, {Shiki}, {Takaki}, {Yoshida}, {Imai}, {Gouda},
  {Gouda}, {Akimoto}, {Honda}, {Hosoya}, {Ikebe}, {Morihana}, {Ohshima},
  {Takagi}, {Takahashi}, {Watanabe}, {Kuroda}, {Morokuma}, {Murata},
  {Nagayama}, {Nogami}, {Oasa}, \& {Sekiguchi}}]{Tanaka2016}
{Tanaka}, Y.~T., {Itoh}, R., {Uemura}, M., {et~al.} 2016, \apj, 823, 35,
  \dodoi{10.3847/0004-637X/823/1/35}

\bibitem[{{Torres} {et~al.}(2020){Torres}, {Casares}, {Jim{\'e}nez-Ibarra},
  {{\'A}lvarez-Hern{\'a}ndez}, {Mu{\~n}oz-Darias}, {Armas Padilla}, {Jonker},
  \& {Heida}}]{Torres2020}
{Torres}, M.~A.~P., {Casares}, J., {Jim{\'e}nez-Ibarra}, F., {et~al.} 2020,
  \apjl, 893, L37, \dodoi{10.3847/2041-8213/ab863a}

\bibitem[{{Vanderburg} \& {Rappaport}(2018)}]{Vanderburg2018}
{Vanderburg}, A., \& {Rappaport}, S.~A. 2018, in Handbook of Exoplanets, ed.
  H.~{Deeg} \& J.~{Belmonte} (Cham: Springer), 2603--2626,
  \dodoi{10.1007/978-3-319-55333-7_37}

\bibitem[{{Vanderburg} {et~al.}(2015){Vanderburg}, {Johnson}, {Rappaport},
  {Bieryla}, {Irwin}, {Lewis}, {Kipping}, {Brown}, {Dufour}, {Ciardi}, {Angus},
  {Schaefer}, {Latham}, {Charbonneau}, {Beichman}, {Eastman}, {McCrady},
  {Wittenmyer}, \& {Wright}}]{Vanderburg2015}
{Vanderburg}, A., {Johnson}, J.~A., {Rappaport}, S., {et~al.} 2015, \nat, 526,
  546, \dodoi{10.1038/nature15527}

\bibitem[{{Veledina} {et~al.}(2019){Veledina}, {Berdyugin}, {Kosenkov},
  {Kajava}, {Tsygankov}, {Piirola}, {Berdyugina}, {Sakanoi}, {Kagitani},
  {Kravtsov}, \& {Poutanen}}]{Veledina2019}
{Veledina}, A., {Berdyugin}, A.~V., {Kosenkov}, I.~A., {et~al.} 2019, \aap,
  623, A75, \dodoi{10.1051/0004-6361/201834140}

\bibitem[{{Wells} {et~al.}(1981){Wells}, {Greisen}, \& {Harten}}]{Wells1981}
{Wells}, D.~C., {Greisen}, E.~W., \& {Harten}, R.~H. 1981, \aaps, 44, 363

\bibitem[{{Wenger} {et~al.}(2000){Wenger}, {Ochsenbein}, {Egret}, {Dubois},
  {Bonnarel}, {Borde}, {Genova}, {Jasniewicz}, {Lalo{\"e}}, {Lesteven}, \&
  {Monier}}]{Wenger2000}
{Wenger}, M., {Ochsenbein}, F., {Egret}, D., {et~al.} 2000, \aaps, 143, 9,
  \dodoi{10.1051/aas:2000332}

\end{thebibliography}
\bibliographystyle{aasjournal}



\end{document}